\documentclass[12pt]{article}
\usepackage{amsfonts,amssymb,amsmath,latexsym}
\newcommand{\be}{\begin{equation}}
\newcommand{\ee}{\end{equation}}
\newcommand{\bea}{\begin{eqnarray}}
\newcommand{\eea}{\end{eqnarray}}

\newcommand{\ba}{\begin{array}}
\newcommand{\ea}{\end{array}}

\newcommand{\bt}{\begin{tabular}}
\newcommand{\et}{\end{tabular}}

\newcommand{\fr}{\frac}
\newcommand{\ci}{\cite}
\newcommand{\cl}{\centerline}
\newcommand{\bs}{\bigskip}

\newcommand{\en}{\eqno}

\newcommand{\bbib}{\begin{thebibliography}}
\newcommand{\ebib}{\end{thebibliography}}

\newcommand{\Llra}{\Longleftrightarrow}

\begin{document}
\titlepage

\centerline{\large TOPOLOGICAL PHASE TRANSITIONS}

\bigskip

\centerline{\large IN LOW-DIMENSIONAL SYSTEMS}

\vspace{1cm}

\centerline{S.A.Bulgadaev\footnotemark{}}\footnotetext{Talk presented
at STATPHYS-20, 19-24 July, 1998, Paris, France

E-mail: bulgad@itp.ac.ru}

\vspace{0.5cm}

\centerline{L.D.Landau Institute for Theoretical Physics}

\centerline{Kosyghin Str.2, Moscow, 117334, RUSSIA}

\bigskip

\centerline{Abstract}

\vspace{0.75cm}

A general theory of the Berezinsky-Kosterlitz-Thouless (BKT) type phase
transitions in low-dimensional systems is proposed. It is shown
that  in d-dimensional case the necessary conditions for it can
take place are 1) conformal invariance of kinetic part of model action and
2) vacuum homotopy group $\pi_{d-1}$  must be nontrivial and
discrete.
It means a discrete vacuum degeneracy for $1d$ systems
and continuous vacuum degeneracy for higher $d$ systems.
For such systems topological exitations have logariphmically
divergent energy and they can be described by corresponding
effective field theories generalizing two-dimensional euclidean
sine-Gordon theory, which is an effective theory of the initial
$XY$-model.
In particular, the effective actions for two-dimensional
chiral models on maximal abelian tori $T_G$ of simple compact groups
$G$ and for one-dimensional models with periodic potentials
are found.
In general case the sufficient conditions
for existence of the BKT type phase transition are 1) constraint
$d \le 2$ and 2) $\pi_{d-1}$ must have some crystallographic symmetries.
Critical properties
of possible low-dimensional
effective theories are
determined and it is shown that in two-dimensional case
they are characterized by the
Coxeter numbers $h_G$ of lattices from the series $\mathbb{A,D,E,Z}$
and can be interpreted as those of conformal field theories
with integer central charge
$c=r,$  where $r$ is a rank of groups $\pi_1$ and $G.$

In one-dimensional case analogous critical properties have
ferromagnetic Dyson chains with discrete Cartan-Ising spins.
In contrast, critical properties of one-dimensional models
with periodic potentials have a weak
dependence on group $G.$

\newpage

A discovery of the possibility of phase transition (PT) in
two-dimensional $XY$-model \cite{sta} from very beginning
attracts a greate attention of theoreticians due to its
unusual properties. First of all it seems that such PT contradicts
to the well-known theorems by Peierls - Landau \cite{pei,lan} and
Bogolyubov - Goldstone \cite{bog,gol} telling us that spontaneous
magnetization and spontaneous  breaking of continuous symmetry
cannot exist in low-dimensional $(d\le 2)$ systems \cite{mer,hoe}. Secondly,
due to the absence of spontaneous magnetization,
correlation functions in low-temperature phase must fall off algebraically
\cite{ric,jan}, what means that the whole low-temperature phase have to
be massless.

All these controversies  were brilliantly resolved
in series of papers by Berezinsky \cite{ber}, Popov \ci{pop}
and by Kosterlitz and Thouless
\cite{kos73,kos74}, who have proven for the first time an important
role of topological
exitations - vortices.
Existence of vortices is connected with the fact that
the manifold of values of $XY$-model
${\cal M}=S^1$ being the simplest  compact manifold with
nontrivial topology  has homotopy group
$\pi_1({\cal M})=\mathbb{Z}.$

Such important role of vortices has reborn an interest
to the topological exitations in quantum field theory,
solid state physics and brings
a discovery of monopoles \cite{pol74,hoo}, instantons \cite{bel,bela},
and others
topological exitations \cite{tou,vol,min}.
But, the main efforts were devoted to the discovery
of topological exitation with {\it finite energy}. All such
exitations give finite contribution to the partition function,
but cannot induce PT similar to the BKT PT, since the latter is induced by
topological exitations with {\it logariphmically divergent energy}.
Instead, topological exitations with finite energy induce a
mass generation through so-called "dimensional
transmutation" mechanism and, consequently, absence of PT in
such models, similar to  all other $\sigma$-models on compact
manifolds \cite{pol}.

Importance of topological exitations with logariphmically
divergent energies for such PT has been earlier discovered
by Anderson, Yuval and Hamann  in their epochal paper \cite{and},
devoted to the Kondo problem. It was shown there that PT
take place in one-dimensional Ising system with long-range $1/r^2$
interaction \cite{dys} due to the presence of logariphmically
interacting domain walls. Though the authors of \cite{kos73} have noted
a similarity between this model and two-dimensional $XY$-model,
but  obvious external differences between these two models (discrete symmetry
and nonlocality of one-dimensional model and continuous symmetry
and locality of two-dimensional model) has not allowed to clear up
this similarity completely.

For more deep understanding of the unifying properties of this two
models it is useful
to note that domain walls are also topological exitations, corresponding
to discrete set of vacuum configurations and consequently can be related
with homotopic group $\pi_0$ of vacuum configuration manifold ${\cal M}$,
which in this case simply means a number of connectivity
components of ${\cal M}$.
Starting from these facts, one can show, that in general case
the condition of existence of
topological exitations with logariphmic
energy puts over the following constraints on non-linear $\sigma$-models:
1) conformal invariance at classical level, 2) their homotopical group
$\pi_{d-1}({\cal M})$ must be nontrivial and discrete.
They  define $\sigma$-models almost uniquely in abitrary dimensions.
Both abovementioned
models satisfy these conditions.

The first property defines form of action $S$
and the second one defines a dimension and form of ${\cal M}$
$$
S=\frac{1}{2\alpha} \int d^d x d^d x' \psi_a (x)\boxtimes_{ab}^{(d)}(x-x')\psi_b (x'),\,
a,b = 1,2...,n,
\eqno(1)
$$
where $\psi \in {\cal M}$ and $n$ is dimension of ${\cal M}$ and
form of kernel $\boxtimes$ depends on dimension of space $d$.
Further, for simplicity, it will be supposed that internal space
is decoupled from
physical space. Than $\boxtimes$ can be decomposed
$$
\boxtimes_{ab}^{(d)}(x)= g_{ab} \Box_d (x),
\eqno(2)
$$
where $g_{ab}$ is, in general, the Euclidean  metric
of the space $\mathbb{R}^{N(n)}$,
in which a manifold ${\cal M}$ can be embedded.
It is more convenient to write $\Box_d$ in the momentum space. For small $k$
$$
\Box_d (k)\simeq |k|^d (1 + a_1 (ka) + ...),
\eqno(3)
$$
where $a$ is a UV cut-off parameter.
Action (1) can be named $d$-dimensional {\emph{conformal}}
nonlinear $\sigma$-model.
The kernel $\Box_d$ generalizes an usual  local and
conformal kernel of two-dimensional $\sigma$-model
$$
\Box(k)\equiv \Box_2(k) = k^2
\eqno(4)
$$
For local models an expression for $\boxtimes$ can be defined in
terms of manifold ${\cal M}$ only
$$
\boxtimes(x) = g_{ab}(\phi)\Box \delta(x)
\eqno(5)
$$
In odd dimensions $\Box_d$ is nonlocal
$$
\Box_d(x)\sim 1/|x|^{2d},
$$
but models with such kernels are used often in physics.

Futher development of the theory of the BKT
PT \cite{jos,sav} shows that in long-wave limit
the partition sum $Z$ of $XY$-model
can be
approximated by product of partition functions of dilute
logariphmic gase (LG) of topological exitations $Z_{LG}$ and
{\emph{free}}
"spin-waves"  $Z_{sw}$
$$
Z_{XY}\simeq Z_{sw} Z_{LG}
\eqno(6)
$$
where  $Z_{LG}$ in its turn
can be represented in the form
of effective  field theory with sine-Gordon  action $S_{SG}$
\cite{jos,wie}. If one assumes that analogous approximate factorization
of partition function takes place for $d$-dimensional conformal
$\sigma$-models with $\pi_{d-1}= \mathbb{Z}$
$$
Z_{\sigma} \simeq Z_{sw}Z_{LG}^{(d)}
\eqno(7)
$$
then $Z_{LG}^{(d)}$ can be represented in the form of effective field
theory with $S_{eff}^{(d)}$
$$
S_{eff}^{(d)} = S_0^d + S_{I} , \,
S_0^d = \frac{1}{2}\int dxdx' \phi(x)\widehat{\Box}_d (x-x')\phi(x')
\eqno(8)
$$
$$
S_{I}= \int dx V(\phi), \, V(\phi) = \mu^2 Cos(\beta \phi)
$$
where an operator $\widehat{\Box}_d$ is inverse to the logariphmic
interaction function, which can be equalized to free correlation function
$$
D(x)= <\phi(x) \phi(0)> =
D(x)-D(0) \sim -\frac{1}{2\pi} \log |x/a|,
\eqno(9)
$$
and for this reason its Fourier components at small $k$ have the form
$$
\widehat{\Box}_d(k) \simeq B_{d}|k|^d(1 + b_1(ak) + ...).
\eqno(10)
$$
It follows from (10) that form of $\widehat{\Box}_d$ coinsides
(up to the coefficients) with that of
{\emph {conformal invariant}}
kernel $\Box_d.$
As a result one gets that conformal $\sigma$-models on compact
spaces with $\pi_{d-1}=\mathbb{Z}$
in long-wave limit are equivalent to non-compact field theories
with effective action, free part of which
has the same form as initial one
and potential term  has a corresponding periodicity. The latter theories
can be named linear conformal $\sigma$-models, associated with conformal
non-linear $\sigma$-models.
This approximate equivalence can be considered as some kind of
{\emph {duality}}
relation between compact conformal and noncompact theories.

Similar representation takes place not only for models with
$\pi_{d-1}$ of rank
$ r(\pi_{d-1}) \equiv r = 1,$
but also for models with $r>1$ and with $\pi_{d-1}$ not equal
to the direct sum
$\pi_{d-1} \ne \bigoplus_1^r \mathbb Z_i.$

In one-dimensional case there are such models with
finite $\pi_0$
and with infinite $\pi_0 = \mathbb L$, where $\mathbb L$ is some
$r$-dimensional lattice.
To the first type belong
the Dyson spin chains with
Cartan-Ising spins \cite{car,bul84}, generalizing Ising spins, which are
the weights of the fundamental representation of group $SU(2)$,
on the Cartan weights of the irreducible representations of other simple
compact groups $G$. The latter form the discrete sets
of classical spins $\{{\bf s}_a \}, \, a=1,...,q,$ where $q$
is dimension of representation,
invariant under some point symmetry groups, so called Weyl groups $W_G.$
For example, $q$-state Potts model spins are a particular
case of weights of fundamental
representation of $G=SU(q)$. In these cases the elements of the
corresponding $\pi_0$
are simply  spin states. These spin chains correspond to the conformal
nonlinear $\sigma$-models.
To the second type belong  conformal
non-compact linear
$\sigma$-models with periodic potentials. Under abovementioned duality
transformation they transform into non-compact models with periodicity
of the dual lattice \ci{sch,bul3}.

In two-dimensional case they include models of crystall melting \ci{nel,nela}
and $\sigma$-models
on maximal abelian tori $T_G$ of the simple compact groups $G,$
generalizing $XY$-model,  with homotopical group
$\pi_1(T_G) = \mathbb{L}_{r^v}$ (for simply connected $G$),
where $\mathbb{L}_{r^v}$ is a
dual root lattice
of the Lie algebra $\mathfrak{G}$ of the group $G$
\cite{bul2}. The corresponding topological exitations - vortices -
have isovectorial topological
charges ${\bf Q} \in \mathbb{L}_{r^v},$ interacting through logariphmic law
$$
\sim ({\bf Q}_1 {\bf Q}_2)\ln |x_1-x_2/a|
$$

And in three-dimensional case they can be conformal (or van der Vaals,
since $\Box_3(x) \sim 1/|x|^6$)
$\sigma$-models on the maximal flag spaces $F_G=G/T_G$ of the simple compact
groups $G$, with $\pi_2(F_G)= \mathbb{L}_{r^v}$ \cite{bul2}.
Flag spaces include as a
particular case sphere $S^2 = SU(2)/U(1)$. The corresponding
topological exitations - instantons - also have isovectorial
topological charges ${\bf Q} \in \mathbb{L}_{r^v}$ and  logariphmic energy.

In all these cases an effective noncompact field theory will have
a potential of the form
$$
V(\vec \phi) = \mu^2 \sum_{\{{\bf r}^v_a \}} \exp(i\beta({\bf r}^v_a \vec \phi))
\eqno(11)
$$
where $\{{\bf r}^v_a\}$ is a set of the minimal dual roots of the Lie
algebra $\mathfrak{G},$
which characterize minimal topological charges of the
possible topological exitations with logariphmic energies.
All these roots can belong only to four different root lattices
$\mathbb{L}_G$ connected with corresponding groups $G$:
$\mathbb{A}_n \Llra G = SU(n+1), \mathbb{D}_n \Llra G = O(2n),
\mathbb{E}_n \Llra G = E_n (n=6,7,8), \mathbb{Z}^n \Llra G = O(2n+1).$
To the exceptional groups $G_2,F_4$ correspond lattices
$\mathbb{A}_2$ and $\mathbb{D}_4$ respectively.
The lattices $\mathbb{L}_G$ are invariant under corresponding affine or
crystallographic symmetry groups ${\cal E}_l = W_G \rtimes \mathbb{L}_G.$

The BKT type PT can be investigated  by renormalization of
the corresponding effective field theories \cite{wie,ami,nel,bul1,bul2}.
One can show that, in general,
PT of BKT type can take place only for $d \le 2,$ since conformal
symmetry in $d>2$ is finite-dimensional and is broken by renormalization
\ci{bul1}. For this reason the following discussion will be
concentrated on models in space with $d\le 2$, where conformal group is
infinite-dimensional. Just the effective field theories connected
with lattices $\mathbb{L}_G$ are renormalizable \ci{bul1}.

As is well known critical singularities at BKT type PT are essential
instead of algebraic ones at II order PT. For example, a
correlation length $\xi$
$$
\xi \sim a\exp (A \tau^{-\nu}),\, \tau = \frac{T-T_c}{T_c}
$$
where $a$ is some UV cut-off parameter and $A$ is a
nonuniversal constant $\sim O(1).$

It appears that in two-dimensional case all critical properties
of  models with $\pi_1 = \mathbb{L}_{r^v}$ are determined
by the Coxeter numbers
$h_G$ of the corresponding lattices $\mathbb{L}_{r^v}$ and groups $G$
$$
h_G = \frac{\mbox{(number of roots)}}{\mbox{(rank of group)}}
$$
For groups from series $A,D,E$ $h_G$ is connected with
second Casimir operator in adjoint representation $K_2= 2h_G.$

In particular, it was found that the critical exponent
$$
\nu_G = 2/(2+h_G)
\eqno(12)
$$

In Table 1 are deduced all possible values of exponent $\nu_G$

\vspace{.5cm}

\centerline{Table 1}

$$
\ba{|c|c|c|c|c|c|c|c|c|c|}
\hline
{} & {} & {} & {} & {} & {} & {} & {} & {} & {}\cr
G & A_n & B_n & C_n & D_n & G_2 & F_4 & E_6 & E_7 & E_8\cr
{} & {} & {} & {} & {} & {} & {} & {} & {} & {}\cr
\hline
{} & {} & {} & {} & {} & {} & {} & {} & {} & {} \cr
\nu_G & \fr{2}{n+3} & \fr{1}{n} & \fr{1}{2} & \fr{1}{n} & \fr{2}{5}
& \fr{1}{4} & \fr{1}{7} & \fr{1}{10} & \fr{1}{16} \cr
{} & {} & {} & {} & {} & {} & {} & {} & {} & {} \cr
\hline
\ea
$$

\vspace{0.5cm}

The exponents $\nu_{A_1} = \nu_{D_2} = \nu_{B_n}$ correspond to the
initial KT exponent $\nu = 1/2.$  For $V(\vec \phi)$ containing the set
of minimal roots $\{{\bf r}_a\}$ the exponents $\nu_G$  for groups
$B_n$ and $C_n$ pass into one another, since their root sets are
mutually dual. For other groups exponents remain the same.
Since $\nu_G$ depends only on $h_G$, they can coincide
for different groups having different rank and acting in different spaces.
This fact could be important when potential $V(\vec \phi)$ is composed
of the characters of different representations of different groups.

The series $A_n$ possesses the largest set of possible values of $\nu_G,$
since besides exponents of the form $1/k$ (where $k$ are integers
$\ge 2$), it also contains exponents of the form $2/(2k+1)$.

In the low-temperature phase the correlation functions  of the fields
exponentials equal to the free correlation functions with a renomalized
"temperature" $\bar \beta$ which depends on initial values $\beta_0,$
$\bar \beta = lim_{l\to \infty} \beta(l)$ \ci{bul1}:
$$
\left< \prod_{s=1}^{t} exp(i({\bf r}_s \vec \phi(x_s)))\right> =
\prod_{i\ne j}^{t} \left|\fr{x_i-x_j}{a}\right|^{\bar \beta ({\bf r}_i{\bf r}_j)/2\pi},
\; \sum_{i=1}^{t} {\bf r} = 0.
\en(13)
$$
At the PT point (where  $\bar \beta = \beta^{*} = 8\pi/r^2 = 4\pi$) an
additional logariphmic factor, related with the "null charge" behaviour
of $g=(a\mu)^d$ and $\delta=((\beta |r|)^2-4d\pi)/4d\pi$ on the
critical separatrix --
the phase separation line,
appears in them:
$$
\prod_{i\ne j}^{t} \left(\ln \left|\fr{x_i-x_j}{a}\right|
\right)^{\beta^{*}({\bf r}_i{\bf r}_j)/2\pi A_G}=
\prod_{i\ne j}^{t} \left(\ln \left|\fr{x_i-x_j}{a}\right|
\right)^{h_G \cos({\bf r}_i{\bf r}_j)},
\en(14)
$$
where $A_G= 4/h_G$ is a coefficient in RG equations for $\delta$ on
the critical separatrix (schematic phase diagram see on Fig.1)

\bs

\begin{picture}(400,150)(-75,0)
\put(100,0){\vector(1,0){100}}
\put(100,0){\vector(0,1){100}}
\put(100,0){\line(-1,0){100}}
\put(100,0){\line(2,1){100}}
\put(170,20){low T phase}
\put(58,60){massive phase}
\put(210,0){$\delta$}
\put(100,110){g}
\put(100,-15){0}
\end{picture}

\vspace{1cm}

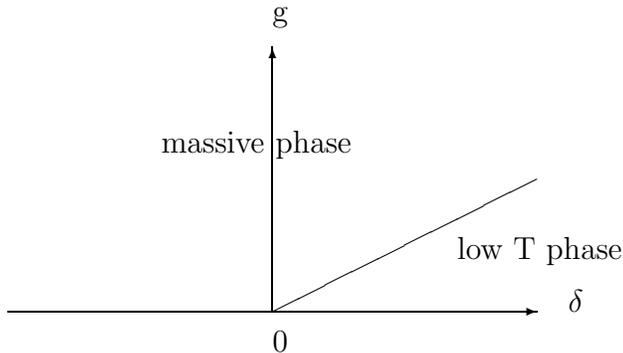
\cl{ Fig.1. Schematic phase diagram of two-dimensional models.}

\vspace{0.5cm}

Free-like behaviour of correlation functions in low-temperature phase
(except logariphmic corrections)
gives one a possibility to use for their description conformal
field theories with integer central charges $C = r$, like $C=1$ theories
in \ci{zam} and instead of
PT points of two-dimensional systems with discrete symmetries, which
are described by conformal theories with rational central charges
\ci{belb,fri,dot,anr,hus}. The BKT type PT can be considered as some
degeneration of II order PT. In this relation it is interesting that
$\nu_G$ coincides with "screening" factor in formulas for central charges
of affine Lie algebras $\hat {\mathfrak{G}}$ at level $k=2$ \ci{kac}
$$
C_k = \frac{k}{k+h_G} dim G
$$
and of coset realization of
minimal unitary conformal models at level $k=1$ \ci{god}
$$
C_k = r\left(1-\frac{h_G(h_G +1)}{(k+h_G)(k+h_G +1)}\right).
$$

\bs

In one-dimensional case a situation is different. There are two
possibilities. For models with finite $\pi_0({\cal M})$ (they include
ferromagnetic Dyson chains with Cartan-Ising spins \cite{car,bul84},
double-well systems with dissipation \cite{bra,leg})
critical exponents $\nu$ coincide with corresponding two-dimensional
ones \cite{car,bul84}.

For non-compact models in periodic potential with infinite discrete
$\pi_0 ({\cal M}) = \mathbb{L}_i, \, i= \mathbb{A,D,E,Z}$ critical exponents
$\nu$ of the essential singularity
weakly depend on type of lattice due to non-renormalization
of the kernel $\Box_1$ and can take only one value $\nu=1$
for all lattices \cite{bul1,bul3}.
Their schematic phase diagrams are depicted on Fig.2.

\begin{picture}(400,150)(-80,0)
\qbezier(100,0)(115,10)(120,100)
\put(90,30){1}
\put(120,30){2}
\put(0,0){\vector(1,0){200}}
\put(0,0){\line(0,1){100}}
\put(0,100){\vector(1,0){200}}
\put(100,0){\line(0,1){100}}
\put(130,60){low T phase}
\put(15,60){massive phase}
\put(210,0){$\delta$}
\put(-10,50){g}
\put(0,-10){0}
\put(100,-10){0}
\put(0,110){$\infty$}
\end{picture}

\vspace{1cm}

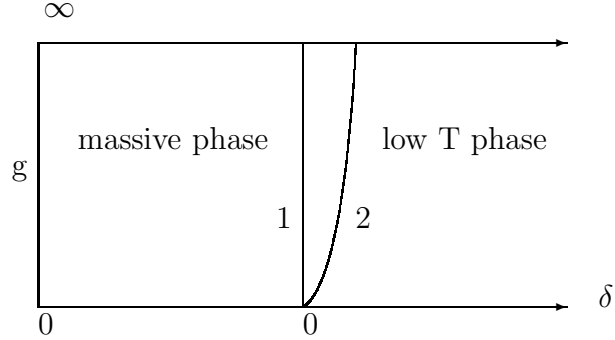
\cl{ Fig.2. Schematic phase diagram of one-dimensional non-compact models:}
\cl{1 - models with $\pi_0 = \mathbb{Z}$;
2 - models with $\pi_0 = \mathbb{A,D,E}.$}

\vspace{1cm}

As a consequence of absence of the kernel renormalization the correlation
functions in low-temperature phase do not have logariphmic corrections
\ci{bul1}.

Last time have appeared many low-dimensional models where topological PT
can take place. Effective field theories discussed above can be applied
for their description due to their universality.
For example, two-dimensional models on $T_G$ can be applied
for discussion of properties of strings, compactified on $T_G$ \ci{gre}
(see case of $S^1$ in \ci{kog,kle} and
of the "decompactification" transitions in perturbed chiral models
on compact groups \ci{bul96}.

One-dimensional  models are applied for description of many
solid state systems. Among them are quantum
macroscopic systems with "ohmic" dissipation \ci{cal,lar,amb,bra,cha,leg,
sch,bul3,scho},
quantum wires \ci{kan}, tunneling between edge states in systems
with quantum Hall effect \ci{moo}, single electron boxes \ci{fal} and
many others. Their number is constantly growing.

This paper is supported by RFBR grants 96-02-17331-a, 96-1596861.

\begin{thebibliography}{99}
\bibitem{ami} Amit D.J., Goldschmidt Y.Y., Grinstein G.,
J.Phys. {\bf A13} (1980) 585.
\bibitem{amb} Ambegaokar V., Eckern U., Schon G., Phys.Rev.Lett.
{\bf 48} (1982) 1745.
\bibitem{and}  Anderson P.W., Yuval G.,Hamann D.R.,
Phys.Rev. {\bf B1} (1970) 4464.
\bibitem{anr}  Andrews G.E., Baxter R.J., Forrester P.J.,
J.Stat.Phys. {\bf 35} (1984) 193.
\bibitem{bel}  Belavin A.A., Polyakov A.M., Pisma v ZETP {\bf 22} (1975) 245.
\bibitem{bela}   Belavin A.A., Polyakov A.M., Schwartz A.S.,
Tyupkin Yu.S. {\bf 59B} (1975) 85.

\bibitem{belb}   Belavin A.A., Polyakov A.M., Zamolodchikov A.B.,
    Nucl.Phys. {\bf B241} (1984) 333.

\bibitem{ber}  Berezinsky V.L., ZETP {\bf 59} (1970) 907,
{\bf 61} (1971) 1144.
\bibitem{bog} Bogolyubov N.N., Selected works, v.{\bf 3}, Kiev,
Naukova dumka, 1971.
\bibitem{bra}  Bray A.J.,Moore M.A., Phys.Rev.Lett. {\bf 49} (1982) 1545.
\bibitem{bul1}  Bulgadaev S.A., Phys.Lett. {\bf 87B} (1979) 47;
Phys.Lett. {\bf 86A} (1981) 213;
Teoret.Matem.Fizika {\bf 49} (1981) 7;
ibid. {\bf 51} (1982) 424.
\bibitem{bul2}  Bulgadaev S.A., Nucl.Phys. {\bf B224} (1983) 349;
Pisma v ZETP {\bf 63} (1996) 743;
ibid. {\bf 63} (1996) 758.

\bibitem{bul84}  Bulgadaev S.A., Phys.Lett. {\bf 102A} (1984) 260.

\bibitem{bul3}  Bulgadaev S.A., Pisma v ZETP {\bf 39} (1984) 264;
ZETP {\bf 90} (1986) 634.
\bibitem{bul96}  Bulgadaev S.A., On decompactification
transition in two-
    dimensional $\sigma$-models. Talk given at
    International conference
    on "Conformal Field Theories and Integrable Models",
    Chernogolovka, Russia, 24-29 June, 1996. Extended version
    Landau Institute preprint 02/06/97, 1997.
\bibitem{cal}  Caldeira A.O., Leggett A.J.,
Phys.Rev.Lett. {\bf 46} (1981) 211,
    Ann.Phys. {\bf 149} (1983) 374.
\bibitem{car}  Cardy J.L., J.Phys. {\bf 14A} (1981) 1407.
\bibitem{cha}  Chakravarty S., Phys.Rev.Lett. {\bf 49} (1982) 681.
\bibitem{dot}  Dotsenko Vl.S.,Fateev V.A.,
Nucl.Phys.{\bf B240} (1984) 312, {\bf B251} (1985) 691.
\bibitem{dys}  Dyson F., Comm.Math.Phys. {\bf 12} (1969) 91, 212;
{\bf 21} (1971) 269.
\bibitem{fal}  Falci G.,Schon G.,Zimany G.T, Phys.Rev.Lett.
{\bf 74} (1995) 3257.
\bibitem{fen}  Fendley P., Ludwig A.,Saleur H.,
Phys.Rev.Lett. {\bf 74} (1995) 3005.
\bibitem{fri}   Friedan D., Qiu Z., Shenker,
Phys.Rev.Lett. {\bf 53} (1984) 1575.
\bibitem{god} Goddard P., Kent A., Olive D., Phys.Lett. {\bf B152}(1985) 88,
Commun.Math.Phys. {\bf 103} (1986) 105.
\bibitem{gol} Goldstone J., Nuovo Cim. {\bf 19} (1961) 154.

\bibitem{gre}   Green M., Schwarz J.H., Witten E., Superstrings Theory,
Cambridge, 1988, vol.{\bf 1,2}.
\bibitem{hoe}  Hohenberg P.C., Phys.Rev. {\bf 158} (1967) 383.
\bibitem{hoo}  t'Hooft G., Nucl.Phys.{\bf B79} (1974) 276.
\bibitem{hus}  Huse D.A., Phys.Rev. {\bf B30} (1984) 3908.
\bibitem{jan} Jancovici B., Phys.Rev.Lett. {\bf 19} (1967) 20.
\bibitem{jos}  Jose J.,Kadanoff L., Kirkpatrick S.,Nelson D.,
Phys.Rev. {\bf B16} (1977) 1217.
\bibitem{kac} Kac V.N., Infinite dimensional Lie algebras, Cambridge
University Press, 1990.
\bibitem{kan}  Kane C.L.,Fisher P.R., Phys.Rev. {\bf B46} (1992) 15233.

\bibitem{kle}   Klebanov I., in String Theory and Quantum Gravity'91,
 Proceedings of the Trieste Spring School,  Workshop ICTP, Trieste, Italy,
World Scientific, 1991.
\bibitem{kog}  Kogan Ya.I., Pisma v ZETP, {\bf 45} (1987) 556.

\bibitem{kos73}  Kosterlitz J.M.,Thouless J.P., J.Phys. {\bf C6} (1973) 118.
\bibitem{kos74}  Kosterlitz J.M., J.Phys. {\bf C7} (1974) 1046.

\bibitem{lan} Landau L.D., ZETP {\bf 7} (1937) 627.
\bibitem{lar}  Larkin A.I., Ovchinnikov Y.N., ZETP {\bf 85} (1983) 1510.
\bibitem{leg}  Legget A.J. et al. Rev.Mod.Phys. {\bf 59} (1987) 1.

\bibitem{min} Mineev V.P., Volovik G.E.,Phys.Rev. {\bf B18} (1978) 3197.
\bibitem{mer}  Mermin N., Wagner H., Phys.Rev.Lett., {\bf 17} (1966) 1133.
\bibitem{moo} Moon K., Yi H., Kane C.L., Girvin S.M., Fisher M.P.A.,
Phys.Rev.Lett. {\bf 71} (1993) 4381.

\bibitem{nel} Nelson D.R., Phys.Rev. {\bf B18} (1978) 2318;
\bibitem{nela} Nelson D.R., B.I.Halperin, Phys.Rev.{\bf B19} (1979) 2457.
\bibitem{oht} Ohta T., Prog.Theor.Phys. {\bf 60} (1978) 968.
\bibitem{pei}  Peierls R.E., Ann.Inst.Henry Poincare {\bf 5} (1935) 177.

\bibitem{pol74}  Polyakov A.M., Pisma v ZETP {\bf 20} (1974) 430.
\bibitem{pol}  Polyakov A.M., Gauge Fields and Strings, Harwood Academic
Publishers, 1987.
\bibitem{pop}  Popov V.N., Feynman integrals in quantum field theory
and statistical mechanics, Moscow. Atomizdat. 1976.

\bibitem{ric} Rice T.M., Phys.Rev.{\bf 140} (1965) 1889.
\bibitem{sav}  Savit R., Phys.Rev.B17 (1978) 1340-1350.
\bibitem{sch}  Schmid A., Phys.Rev.Lett. {\bf 51} (1983) 1506.
\bibitem{scho}  Schon G.,Zaikin A.D., Phys.Rep.{\bf 198} (1990) 237.

\bibitem{sta}  Stanley H.E., Kaplan T.A., Phys.Rev.Lett.{\bf 17} (1966) 913.
\bibitem{tou} Toulouse G., Kleman M., J.Physique Lett. {\bf 37} (1976) 149.
\bibitem{vol} Volovik G.E., Mineev V.P., ZETP {\bf 45} (1977) 1186.
\bibitem{wie} Wiegmann P.B., J.Phys.{\bf C11} (1978) 1583.
\bibitem{zam}  Zamolodchikov A.B., Zamolodchikov Al.B.,
Sov.Sci.Rev. {\bf A10} (1989) 271.
\ebib

\end{document}